# Weighted Sum Power Minimization for Cooperative Spectrum Sharing in Cognitive Radio Networks

Yang Yu*
yuyang@tgc.edu.cn
Hubei Three Gorges Polytechnic

*Abstract*—This letter introduces weighted sum power (WSP), a new performance metric for wireless resource allocation during cooperative spectrum sharing in cognitive radio networks, where the primary and secondary nodes have different priorities and quality of service (QoS) requirements. Compared to using energy efficiency (EE) and weighted sum energy efficiency (WSEE) as performance metrics and optimization objectives of wireless resource allocation towards green communication, the linear character of WSP can reduce the complexity of optimization problems. Meanwhile, the weights assigned to different nodes are beneficial for managing their power budget. Using WSP as the optimization objective, a suboptimal resource allocation scheme is proposed, leveraging linear programming and Newton's method. Simulations verify that the proposed scheme provides near-optimal performance with low computation time. Furthermore, the initial approximate value selection in Newton's method is also optimized to accelerate the proposed scheme.

## I. INTRODUCTION

COGNITIVE radio, profoundly impacting technologies such as device-to-device (D2D) communication in the 4G and 5G eras, allows secondary users (SUs) to share primary users' (PUs) authorized spectrum via overlay or underlay spectrum sharing. In addition to these two typical modes, cooperative spectrum sharing, which enables an SU to access the authorized spectrum by acting as a relay for the corresponding PU, has been widely investigated in the past decade [1]. This paradigm guarantees the PU's QoS while providing transmission opportunities for the SU, resulting in a win-win situation.

Recently, green communication has become a primary trend in structuring communication networks due to concerns such as the economy, environment, health, and sustainable development. Energy efficiency (EE), i.e., the system throughput to the power consumption ratio, has been accepted as one of the most meaningful metrics to judge the system benefit-cost ratio. Nevertheless, more information about the priorities of different nodes, especially for heterogeneous networks, needs to be reflected. On the other hand, in cooperative spectrum sharing, regarding overall EE as the objective for resource allocation will lead to a fractional programming (FP) problem, which is hard to tackle [2]. To avoid the first limitation, weighed sum energy efficiency (WSEE) has been proposed as an alternative metric that balances global system performance and priorities of different nodes. Nevertheless, WSEE belongs to the family of multiple-ratio FP problems and is still challenging to solve.

The authors of [3] focused on minimizing system energy consumption in cooperative spectrum sharing and proposed a scheme based on Karush-Kuhn-Tucker (KKT) conditions. Inspired by this idea, we introduce weighted sum power (WSP) as an alternative metric to WSEE. WSP is a linear weighted sum of power consumption, and its minimization can usually be converted into a convex form. Moreover, since WSP is directly linked to the power consumption of different nodes, the power budgeting of different nodes can be managed by dynamically adjusting corresponding weights, thereby extending the network lifetime, which is significant for energy-limited wireless networks, e.g., low-power wide area networks (LPWAN). Existing works have extensively studied resource allocation for cooperative spectrum sharing. Besides [3], most network-centric schemes utilized the Lagrange multiplier method and KKT conditions. For example, [4] proposed a WSEE maximization scheme for public-interest D2D users based on spectrum-power trading. [5] and [6] proposed non-orthogonal multiple access (NOMA) assisted schemes with simultaneous wireless information and power transfer (SWIPT) to optimize the sum rate. Various mathematical frameworks, including contract and game theories, have also been applied from the user-centric perspective. For example, [1] investigated the optimal contract design under incomplete information. [7] further investigated the incentive mechanism in orthogonal frequency-division multiple access (OFDM)-based cognitive Internet of Things (IoT) networks. In [8], the authors introduced a two-timescale scheme to improve spectrum efficiency (SE) using matching game theory. However, some of these works made additional assumptions to simplify their problems, e.g., SUs' transmit power of different stages in [8] is equal. More importantly, optimizing resource allocation with WSP as the objective has yet to be studied. Even the work that inspired us, i.e., [3], mistakenly considered resource allocation a convex problem.

In this letter, we propose a resource allocation scheme by taking WSP as the optimization objective for cooperative spectrum sharing. The coupled power and spectrum allocation is decomposed into a standard linear programming problem with two variables and a problem of searching the extremum of a univariate function. We solve these two subproblems via the graphic and Newton's method. Intrinsically, the proposed scheme is suboptimal but provides near-optimal performance with low computational complexity. In addition, the initial value selection in Newton's method is also optimized, further improving the feasibility of the proposed scheme for real-time applications.

*Corresponding author.

## II. SYSTEM MODEL AND PROBLEM FORMULATION

As mentioned in Section I, WSP minimization belongs to weighed-sum optimization problems. Thus, from the network-centric perspective, we can break the resource allocation between $M$ primary and $N$ secondary links, an integer programming problem, down into $MN$ subproblems for the potential matched link pairs and a minimum weighted bipartite graph matching problem. The latter can be solved through the classic Kuhn-Munkres algorithm. Based on this analysis, we consider a pair of matched primary and secondary links. Fig. 1 depicts a sample sketch of the system model and frame structure as an application for D2D communication in OFDMA-based cellular networks. This system supports full-duplex communications. However, since the base station (PR) has relatively high transmit power and a sufficient energy budget, cooperative spectrum sharing is rarely needed for downlink transmission. Thus, due to the limited energy budget for mobile terminals, we focus on uplink transmission and assume the link pair operates over a single physical resource block (PRB). Each frame starts with the preprocessing phase, followed by $T$ cooperative spectrum-sharing subframes.

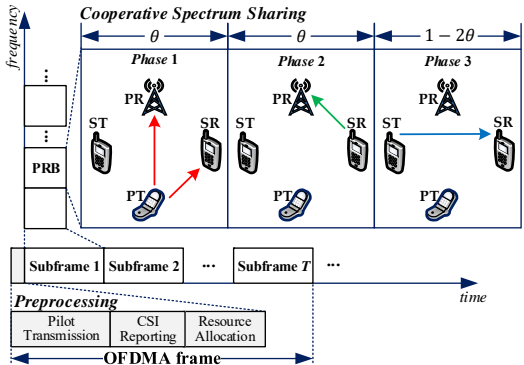
Fig. 1. An illustration of the system model and frame structure.

Assume each subframe involves three phases [3], [8]. In the first phase, the primary transmitter (PT) broadcasts with power $p_p$ to the primary receiver (PR) and the SU, which acts as the relay. Since the choice of the secondary transmitter (ST) or receiver (SR) barely impacts the problem formulation and solution, we take the SR as the relay. Moreover, while either the amplify-and-forward (AF) protocol or the decode-and-forward (DF) protocol has its relative merits, this paper takes the DF relay strategy as an example [3], [8]. In phase two, the SR relays the data of the primary link with power $p_r$. In response to the assistance from the SR, the primary link allows the ST to transmit data with power $p_s$ in phase three. The first two phases for the relay transmission of the primary link last for $\theta \in (0, 0.5)$ subframe, respectively, and phase three lasts for $1 - 2\theta$ subframe [3], [8]. Accordingly, $\theta p_p$, $\theta p_r$, and $(1 - 2\theta)p_s$ are the power consumptions of the PT, SR, and ST during one subframe. If we use $w_{pt}$, $w_{sr}$, and $w_{st}$ to represent the weights of the PT, ST, and SR, the system WSP is given by the following expression:

$$\mathcal{W} = w_{pt}\theta p_p + w_{sr}\theta p_r + w_{st}(1-2\theta)p_s, \quad (1)$$

where $u = w_{pt}\theta p_p + w_{sr}\theta p_r$ and $v = w_{st}(1-2\theta)p_s$ are the components contributed by the primary and secondary links, respectively.

Due to hardware constraints in practical systems such as LTE, this letter considers discrete power levels and assumes that $p_{min}$ and $p_{max}$ are the minimum and maximum transmit powers, respectively. Similarly, the spectrum allocation factor $\theta$ is assumed to be discrete.

Furthermore, to guarantee the two links' quality of service (QoS), their achievable SEs must stay above certain thresholds. Herein, we denote $g_{pp}$, $g_{ps}$, $g_{sp}$, and $g_{ss}$ as channel gains between the PT and the PR, the PT and the SR, the SR and the PR, the ST and the SR, respectively. With the noise power density $n_0$, we can define the auxiliary parameter $\lambda_{pp} \triangleq g_{pp}/n_0$, $\lambda_{ps} \triangleq g_{ps}/n_0$, $\lambda_{sp} = g_{sp}/n_0$, and $\lambda_{ss} = g_{ss}/n_0$. Then, according to the Shannon-Hartley theorem and referring to [3], [8], we can express the SEs of the primary and secondary links as

$$S_p = \theta \log_2(1 + \min\{\lambda_{ps}p_p, \lambda_{pp}p_p + \lambda_{sp}p_r\}) \geq Q_p, \quad (2)$$
$$S_s = (1-2\theta)\log_2(1 + \lambda_{ss}p_s) \geq Q_s, \quad (3)$$

where $Q_p$ and $Q_s$ denote the minimum SE requirements.

From (2) and (3), we can obtain

$$p_p^{low} = max\{p_{min}, (2^{Q_p/\theta}-1)/\lambda_{ps}\}, \quad (4)$$
$$p_s^{low} = max\{p_{min}, [2^{Q_s/(1-2\theta)}-1]/\lambda_{ss}\}, \quad (5)$$

as two lower bounds on $p_p$ and $p_s$.

Thus, we can formulate the WSP minimization problem as

$$\min_{p_p, p_r, p_s, \theta} \mathcal{W}, \quad (6)$$
$$\text{s.t.} \quad 0 < \theta < 0.5, \quad (6a)$$
$$\lambda_{pp}p_p + \lambda_{sp}p_r \geq 2^{Q_p/\theta} - 1, \quad (6b)$$
$$p_p^{low} \leq p_p \leq p_{max}, \quad (6c)$$
$$p_{min} \leq p_r \leq p_{max}, \quad (6d)$$
$$p_s^{low} \leq p_s \leq p_{max}, \quad (6e)$$

where (6b) is derived from (2). Contrary to the conclusion in [3], (6) is a nonconvex problem, which can be revealed from the Hessian matrix of $\mathcal{W}$. We calculate its eigenvalues as 0, 0, $\sqrt{w_{pt}^2 + 4w_{st}^2 + w_{sr}^2}$, and $-\sqrt{w_{pt}^2 + 4w_{st}^2 + w_{sr}^2}$. Thus, it is not a positive semi-definite matrix, and $\mathcal{W}$ is nonconvex [9]. Due to the above analysis, the authors' optimization method in [3] is incorrect and cannot be followed.

## III. WSP MINIMIZATION-BASED RESOURCE ALLOCATION

Since (6) is nonconvex, we cannot apply Slater's condition to prove the strong duality between it and its Lagrangian dual problem. Moreover, it can be verified that the duality gap is not zero. In this situation, the KKT conditions cannot provide necessary and sufficient conditions for optimality [9]. If we apply them to (6), there will be $2^L$ cases that need to be discussed according to $L$, the number of Lagrange multipliers. Therefore, it is challenging to derive a computationally efficient algorithm and achieve the global optimal minimum of (6) by exploiting the KKT conditions. As an alternative, we leverage the graphic and Newton's method in the sequel to explore an efficient suboptimal solution. First, with a fixed $\theta$, it can be easily observed that $\mathcal{W}$ is monotonically increasing with $p_s$. Thus, we can get the solution $p_s^*$ as $p_s^{low}$. Consequently, $p_p^*$ and $p_r^*$ can be obtained via the following problem.

$$\min_{p_p, p_r} u/\theta = w_{pt} p_p + w_{sr} p_r, \qquad (7)$$
$$\text{s.t. (6b), (6c), and (6d).}$$

(7) is a standard linear programming problem with two variables, which can be visualized on the $xOy$ plane. With isolines $l: u/\theta = w_{pt} x + w_{sr} y$ and $m: 2^{Q_p/\theta} - 1 = \lambda_{pp} x + \lambda_{sp} y$, we demonstrate (7) in Fig. 2, where points A, B, C, and D are located at $(p_{max}, p_{max})$, $(p_p^{low}, p_{max})$, $(p_p^{low}, p_{min})$, and $(p_{max}, p_{min})$, respectively.

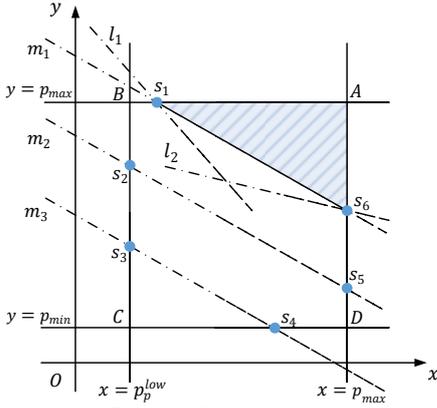

Fig. 2. An illustration of (7) in the $xOy$ plane.

For any $\theta$, the shaded area denotes $S$, the feasible set of (7). Noting that $u$ will decrease when $l$ moves down, the solution $(p_p^*, p_r^*)$ is at the first intersection of $l$ and $S$. In order to prevent $S$ from being $\emptyset$, $f(p_{max}, p_{max}) \geq Q_p$, $p_p^{low} \leq p_{max}$, and $p_s^{low} \leq p_{max}$ must be satisfied, where $f(x, y) \triangleq \theta \log_2(1 + \lambda_{pp} x + \lambda_{sp} y)$. With the slopes $k_l \triangleq w_{pt}/w_{sr}$, $k_m \triangleq \lambda_{pp}/\lambda_{sp}$ and another two auxiliary functions $g(y) \triangleq \frac{1}{\lambda_{pp}}(2^{Q_p/\theta} - 1 - \lambda_{sp} y)$, and $h(x) \triangleq \frac{1}{\lambda_{sp}}(2^{Q_p/\theta} - 1 - \lambda_{pp} x)$, $(p_p^*, p_r^*)$ can be obtained in the following five cases.

1. If $k_l \geq k_m$ and $f(p_p^{low}, p_{max}) < Q_p \leq f(p_{max}, p_{max})$, we observe that $(p_p^*, p_r^*) = (g(p_{max}), p_{max})$ is on AB, e.g., $s_1$. The first and second derivatives of $u$ can be formulated as

$$u'(\theta) = \frac{w_{pt}}{\lambda_{pp}} \left[ \left(1 - \frac{Q_p \ln 2}{\theta}\right) 2^{Q_p/\theta} - 1 - \lambda_{sp} p_{max} \right] + w_{sr} p_{max}, \quad (8)$$

$$u''(\theta) = \frac{w_{pt}}{\lambda_{pp}} \frac{(Q_p \ln 2)^2}{\theta^3} 2^{Q_p/\theta}. \qquad (9)$$

2. If $k_l \geq k_m$ and $f(p_p^{low}, p_{min}) < Q_p \leq f(p_p^{low}, p_{max})$, it can be seen that the solution lies on BC, e.g., $s_2$ or $s_3$, which equals $(p_p^{low}, h(p_p^{low}))$. Therein, we have

$$u'(\theta) = w_{pt} p_{min} + \frac{w_{sr}}{\lambda_{sp}} \left[ \left(1 - \frac{Q_p \ln 2}{\theta}\right) 2^{Q_p/\theta} - 1 - \lambda_{pp} p_{min} \right], \quad (10)$$

$$u''(\theta) = \frac{w_{sr}}{\lambda_{sp}} \frac{(Q_p \ln 2)^2}{\theta^3} 2^{Q_p/\theta}, \qquad (11)$$

when $p_{min} \geq (2^{Q_p/\theta} - 1)/\lambda_{ps}$. Or

$$u'(\theta) = \frac{w_{sr}}{\lambda_{ps}} \left( k_l - k_m + \frac{\lambda_{ps}}{\lambda_{sp}} \right) \left[ \left(1 - \frac{Q_p \ln 2}{\theta}\right) 2^{Q_p/\theta} - 1 \right], \quad (12)$$

$$u''(\theta) = \frac{w_{sr}}{\lambda_{ps}} \left( k_l - k_m + \frac{\lambda_{ps}}{\lambda_{sp}} \right) \frac{(Q_p \ln 2)^2}{\theta^3} 2^{Q_p/\theta}, \quad (13)$$

when $(2^{Q_p/\theta} - 1)/\lambda_{ps} > p_{min}$.

3. If $k_l < k_m$ and $f(p_p^{low}, p_{min}) < Q_p < f(p_{max}, p_{min})$, $s_4$ denotes $(p_p^*, p_r^*)$, which equals $(g(p_{min}), p_{min})$. There we have

$$u'(\theta) = \frac{w_{pt}}{\lambda_{pp}} \left[ \left(1 - \frac{Q_p \ln 2}{\theta}\right) 2^{Q_p/\theta} - 1 - \lambda_{sp} p_{min} \right] + w_{sr} p_{min}, \quad (14)$$

$$u''(\theta) = \frac{w_{pt}}{\lambda_{pp}} \frac{(Q_p \ln 2)^2}{\theta^3} 2^{Q_p/\theta}. \qquad (15)$$

4. If $k_l < k_m$ and $f(p_{max}, p_{min}) \leq Q_p < f(p_{max}, p_{max})$, $s_5$ or $s_6$ represents $(p_p^*, p_r^*)$, which equals $(p_{max}, h(p_{max}))$ and lies on DA. There are

$$u'(\theta) = w_{pt} p_{max} + \frac{w_{sr}}{\lambda_{sp}} \left[ \left(1 - \frac{Q_p \ln 2}{\theta}\right) 2^{Q_p/\theta} - 1 - \lambda_{pp} p_{max} \right], \quad (16)$$

$$u''(\theta) = \frac{w_{sr}}{\lambda_{sp}} \frac{(Q_p \ln 2)^2}{\theta^3} 2^{Q_p/\theta}. \qquad (17)$$

5. If $f(p_p^{low}, p_{min}) \geq Q_p$, $S$ is the whole ABCD. $(p_p^*, p_r^*)$ is at point C, and we can use (14) and (15) to calculate $u'(\theta)$ and $u''(\theta)$ in this case.

Combining the first and second derivatives of $v$, i.e.,

$$v'(\theta) = \frac{2w_{st}}{\lambda_{ss}} \left[ 1 - \left(1 - \frac{Q_s \ln 2}{1 - 2\theta}\right) 2^{Q_s/(1-2\theta)} \right], \qquad (18)$$

$$v''(\theta) = \frac{4w_{st}}{\lambda_{ss}} \frac{(Q_s \ln 2)^2}{(1-2\theta)^3} 2^{Q_s/(1-2\theta)}, \qquad (19)$$

$\mathcal{W}''(\theta) > 0$ always holds, so $\mathcal{W}'(\theta)$ is strictly increasing. On the other hand, there exists a $\theta'$ that makes $\mathcal{W}'(\theta) = 0$, since $\lim_{\theta \to 0^+} \mathcal{W}'(\theta) = -\infty$ and $\lim_{\theta \to 0.5^-} \mathcal{W}'(\theta) = +\infty$. It is easy to find out that $\mathcal{W}'(\theta) < 0$ holds if $\theta \in (0, \theta')$ and $\mathcal{W}'(\theta) > 0$ if $\theta \in (\theta', 0.5)$. Consequently, $\mathcal{W}$ achieves its minimum at $\theta^* = \theta'$.

To obtain an approximation of $\theta^*$, we adopt an iterative procedure with an initial $\theta_0 \in (0, 0.5)$. A more approximate $\theta_k$ is given via Newton's method as

$$\theta_k = \theta_{k-1} - \mathcal{W}'(\theta_{k-1}) / \mathcal{W}''(\theta_{k-1}), \qquad (20)$$

until $|\theta_k - \theta_{k-1}| \leq \varepsilon$, where $\varepsilon$ is a predefined sufficiently small tolerance. Then, we obtain $\theta^* \approx \theta_k$.

Newton's method provides quadratic convergence with a time complexity of $O(\log \log(1/\varepsilon))$ when $\theta_0$ is sufficiently close to $\theta^*$ in practice. Otherwise, Newton's iteration will lead to a large $k$ or cannot even converge to $\theta^*$. To avoid this, the selection of $\theta_0$ should be improved. For example, we can choose $\theta_0$ to fulfill $\mathcal{W}'''(\theta_0) \neq 0$ and $|\mathcal{W}''(\theta_0)|^2 > |\mathcal{W}(\theta_0) \mathcal{W}'''(\theta_0)|/2$, thus ensuring the convergence of the Newton iteration process in most cases [10]. In addition, we also take $\theta^*$ from the previous subframe as $\theta_0$ in the current subframe. The simulation in Section IV shows that this method especially applies to slowly changing wireless environments, such as quasi-static channels.

Based on the above, we propose a centralized power and spectrum allocation scheme based on WSP minimization for cooperative spectrum sharing. It can be easily inferred that this scheme can potentially extend the network lifetime by dynamically adjusting weights. We will work on optimizing the weights in the future to align with sustainable development.

---

**Algorithm 1:** WSP Minimization-based Resource Allocation

**Choose** $\varepsilon > 0$, and initialize $\theta_0$
**repeat**
  get $\mathcal{U}'(\theta_{k-1})$ and $\mathcal{U}''(\theta_{k-1})$ through the graphic method
  calculate $\theta_k$ using (20)
**until** $|\theta_k - \theta_{k-1}| \le \varepsilon$
Substitute $\theta^* \approx \theta_k$ into the five cases to obtain $(p_p^*, p_r^*, p_s^*)$

---

## IV. NUMRICAL RESULTS

In this section, numerical results are provided to evaluate the performance of the proposed scheme. A pair of matched primary and secondary links is considered, as we explain in Section II. Some of the main simulation parameters are given as follows. $n_0 = -174$ dBm/Hz, $p_{min} = -40$ dBm, $p_{max} = 23$ dBm, and $w_{pt} = w_{st} = w_{sr} = 1$. The granularities of power and spectrum levels are assumed to be $\Delta p = 1$ dB and $\Delta \theta = 0.005$, respectively. We utilize a simplified path loss model with a path loss exponent 3.8.

To demonstrate the benefits of the proposed scheme, we first adopt a Monte Carlo simulation based on 1000 runs. The ST and SR are randomly distributed over a circular area with a radius of 5000 m. At the same time, the PT and PR are located at both ends along the radius direction, respectively. The required minimum SE of the secondary link is set to be $Q_s = 3$ bps/Hz, and $Q_p$ varies from 1 to 5 bps/Hz. The optimal, random, and KKT-based schemes are chosen as baselines. In particular, the optimal scheme is realized via an exhaustive search, and the random scheme randomly selects $p_p, p_r, p_s$, and $\theta$. The KKT-based scheme, often used for network-centric systems, solves (6) by seeking possible local minima of all the $2^L$ subproblems, where $L = 9$ [11]. Most of these subproblems can easily be solved analytically or confirmed as having no solution. The remaining ones can be converted into solving $L_0$ equations as $\mathcal{F}(\theta) = 0$ without analytical solutions. Newton's method or linear search is applied according to the monotonicity of $\mathcal{F}(\theta)$. Therefore, an estimate of the time complexity of the KKT-based scheme is $O(L_0 Q)$. The time complexity of the considered schemes and some existing user-centric schemes mentioned in Section I is summarized in Table I, where $P$ and $Q$ denote the number of power and spectrum levels. $V$ is determined by the QoS requirement and the long-term payoff of the secondary link while $\epsilon = 1/8 \sim 8$.

TABLE I
A SUMMARY OF COMPUTATIONAL COMPLEXITY

| Name | Time complexity |
|---|---|
| Proposed scheme | $O(\log \log(1/\varepsilon))$ |
| Optimal scheme | $O(P^3 Q)$ |
| KKT-based scheme [3], [4], [6] | $O(L_0 Q)$ (an estimate) |
| Contract-based scheme [1] | $O(Q)$ |
| Distributed Matching Algorithm [8] | $O(V/\epsilon)$ |

In Fig. 3, we examine the WSP of each scheme. Since the weights of these nodes are equal, the WSP also represents the system power. We observe that the WSP increases along with $Q_p$. Moreover, the proposed scheme can achieve near-optimal WSP and has a lower WSP than the KKT-based scheme, while the random one shows the highest WSP, which reveals the superiority of our proposed scheme. It is worth noting that a performance gap exists between the optimal and KKT-based schemes since the global minimum in nonconvex optimization may not necessarily satisfy the KKT conditions [11].

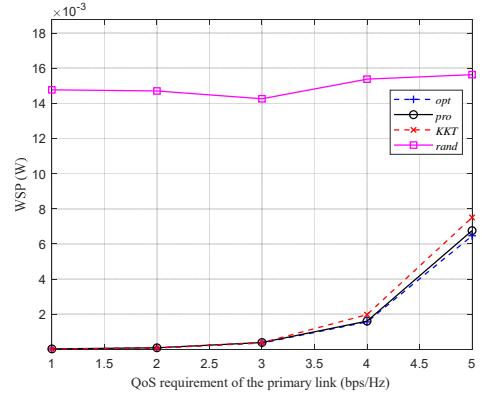

Fig. 3. System WSP versus $Q_p$.

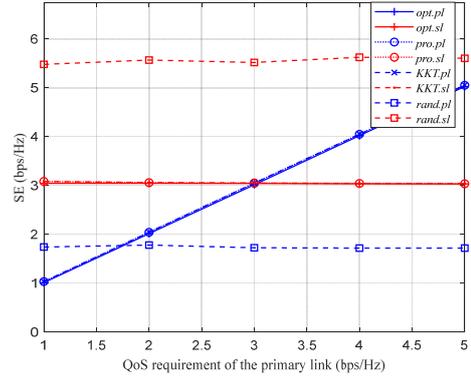

Fig. 4. Achievable SEs of the primary and secondary links versus $Q_p$.

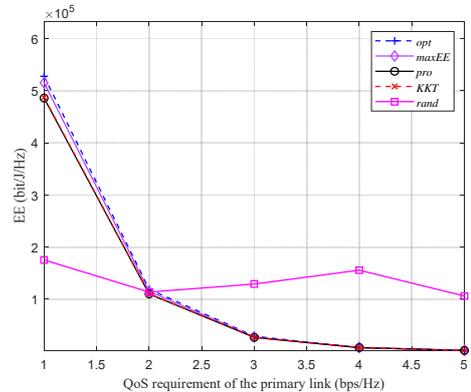

Fig. 5. System EE versus $Q_p$.

Fig. 4 shows that the considered schemes ensure the required minimum QoS of the primary and secondary links except for the random scheme. It is worth noting that the achievable SE of either link is approximately equal to its minimum required QoS in the optimal and proposed schemes. Inspired by this observation and according to the system EE, i.e., $\eta = \frac{S_p + S_s}{\theta p_p + \theta p_r + (1-2\theta)p_s}$, we can use the WSP with $w_{pt} = w_{st} = w_{sr} = 1$ to represent the system EE since $S_p \approx Q_p$ and $S_s \approx Q_s$. Fig. 5 shows the EE versus $Q_p$, where the EE maximization scheme based on an exhaustive search is added to demonstrate this conclusion. We can see the curves of the EE maximization and the optimal scheme overlap. Thus, the WSP minimization scheme with equal weights can replace the EE maximization to a certain extent under certain conditions. However, the latter is more challenging since it is an FP problem [2]. The conditions under which the link SE equals its minimum required QoS are left for future research. We will also try to reveal the relationship between WSP and WSEE in the future.

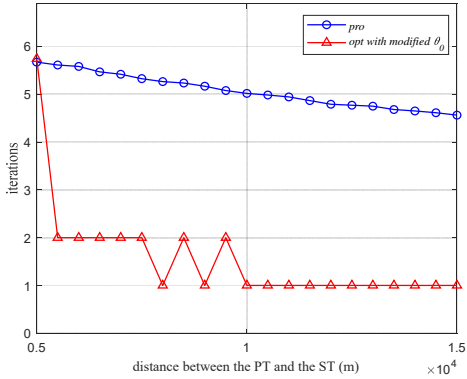

Fig. 6. Iterations of Newton's method versus the distance between the PT and the ST.

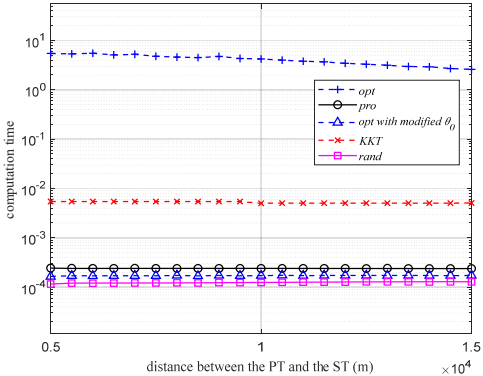

Fig. 7. Computation time versus the distance between the PT and the ST.

Finally, Figs. 6 and 7 show the proposed scheme's convergence speed and computation time when $Q_p = Q_s = 3$ bps/Hz. We set the distances between the PT and PR, the SR and the PR, and the ST and the SR as 10000 m, 5000 m, and 5000 m, respectively. Moreover, the distance between the PT and the SR varies from 5000 to 15000 m. It can be intuitively seen from Fig. 6 that the proposed scheme with a modified $\theta_0$ can reduce the iterations of Newton's method significantly, i.e., from more than 5 times to less than 2, except when the distance between the PT and the SR is equal to 5000 m. Furthermore, in this simulation, we measure the computation time with an AMD Ryzen 9 7945HX CPU with 40 GB memory on MATLAB® R2022b. From Fig. 7, we notice that all the schemes have low computation time except for the optimal one, and the computation time of the KKT-based scheme is about 21 times that of the proposed scheme. In stark contrast, the optimal one's computation time is higher by several orders of magnitude, e.g., more than 5 s, when the distance between the PT and the SR equals 5000 m. Thus, the proposed scheme achieves a good trade-off between system performance and complexity, which suits real-time operation.

## V. Conclusion

In this letter, we have presented the WSP as a performance metric suitable for networks with heterogeneous user priorities and QoS requirements. Based on WSP minimization, we have developed a suboptimal low-complexity power and spectrum allocation scheme for cooperative spectrum sharing. Due to Newton's method, the proposed scheme converges very fast. The numerical results have demonstrated the near optimality of the proposed scheme and highlighted its feasibility for real-time operation. Alongside the proposed scheme, we have modified the initial value selection for Newton iteration. We have also revealed through simulations that the WSP with equal weights can represent the system EE to a certain extent.